# Single Photon Scattering Can Account for the Discrepancies Between Entangled Two-Photon Measurement Techniques

*Bryce P. Hickam[1]†, Manni He[1]†, Nathan Harper[1], Szilard Szoke[2], Scott K. Cushing[1]\**

[1]Department of Chemistry and Chemical Engineering, California Institute of Technology, 1200 E. California Blvd. Pasadena, CA

[2]Department of Engineering and Applied Science, California Institute of Technology, 1200 E. California Blvd. Pasadena, CA

AUTHOR INFORMATION

†These authors contributed equally to this work

**Corresponding Author**

*Corresponding author: scushing@caltech.edu

**ABSTRACT**

Entangled photon pairs are predicted to linearize and increase the efficiency of two-photon absorption, allowing continuous wave laser diodes to drive ultrafast time-resolved spectroscopy and nonlinear processes. Despite a range of theoretical studies and experimental measurements, inconsistencies persist about the value of the entanglement enhanced interaction cross section. A spectrometer is constructed that can temporally and spectrally characterize the entangled photon state before, during, and after any potential two-photon excitation event. For the molecule Rhodamine 6G, which has a virtual state pathway, any entangled two-photon interaction is found to be equal to or lower than classical, single photon scattering events. This result can account for the discrepancies between the wide variety of entangled two-photon absorption cross sections reported from different measurement techniques. The reported instrumentation can unambiguously separate classical and entangled effects and therefore is of importance for the growing field of nonlinear and multiphoton entangled spectroscopy.

**TOC GRAPHICS**

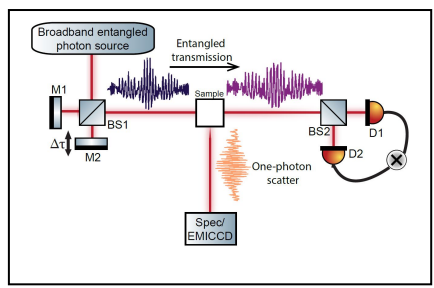

**MAIN TEXT**

It is well known that the inherent correlations between entangled photons lead to spectroscopic enhancements in signal-to-noise[1–4] and diffraction limits[5] as well as measurement techniques such as ghost imaging.[6–12] A new and quickly growing field of entangled two-photon spectroscopy exploits these correlations in a different manner: enhancing excited state and multiphoton interactions.[13–16] For example, early experiments purported that entangled photons can linearize two-photon processes to achieve near one-photon absorption cross sections.[17] This finding would have far-reaching implications in optics and photonics as low power, continuous wave (CW) laser systems could replace pulsed lasers in spectroscopic techniques such as nonlinear bioimaging and ultrafast spectroscopy.[18–20] Combined with nanophotonic platforms that are capable of generating and manipulating entangled photon pairs,[21,22] ultrafast measurements could achieve unprecedented accessibility and availability.

While the entangled photons' linearization of two-photon processes can be derived theoretically,[23–25] intense debate exists over the enhancement factor compared to classical two-photon absorption. Naively, one would expect the entangled two-photon absorption (ETPA) cross section to be close to that of a single photon process, given the linearization of the absorption. However, as shown in Fig. 1, the reported range of ETPA cross sections varies from near single photon levels to non-existent or barely higher than classical two-photon absorption at the same photon flux.[17,19,26–38] Complicating this debate is the disparate measurement techniques based on transmission, coincidence counting, and fluorescence that have been used. An agreed upon measurement process does not exist, and even those that are proposed rely on intensity counting methods instead of temporally and spectrally characterizing the entangled photon state during a proposed interaction event.

In this paper, we construct a spectrometer that temporally and spectrally measures the entangled photon state before, during, and after a proposed entangled two-photon event. The spectrometer is used to measure the proposed ETPA of a prototypical molecular dye, Rhodamine 6G (R6G), which has a well characterized, virtual state-mediated classical two-photon absorption pathway.[39,40] To enhance the chances of measuring a two-photon fluorescence signal, the spectrometer uses a custom, high-flux (20 nW) entangled photon source with a tunable bandwidth and correlation time down to 20 fs.[41] By measuring the temporal and spectral characteristics of the entangled state in transmission, and the 90-degree signal relative to the cuvette, we determine that if any entangled enhancement to the virtual-state mediated two-photon absorption is present, it has a cross section smaller than that of single photon (resonant) scattering ($\sim 10^{-21}$ cm$^2$/molecule).

The conclusion is based on the facts that: 1) the entangled state is nearly identical before and after transmission, with only a slight amplitude loss, and 2) the scattered signal temporally follows the pattern of one-photon interference instead of entangled two-photon interference, with spectral components far from the fluorescence spectrum of the dye. The single photon scattering explains the past discrepancies in measured cross sections between different techniques. Moreover, the paper presents a blueprint for a spectrometer that can unambiguously determine entangled photon excited state effects in the rapidly growing field of entangled nonlinear and multiphoton spectroscopy. The scattering limited cross section for R6G also bolsters current theories that suggest virtual state pathways have cross sections that are beyond noise limits of single photon detectors and that real intermediate states may be necessary for entangled two-photon excitation events with high cross sections.[42,43]

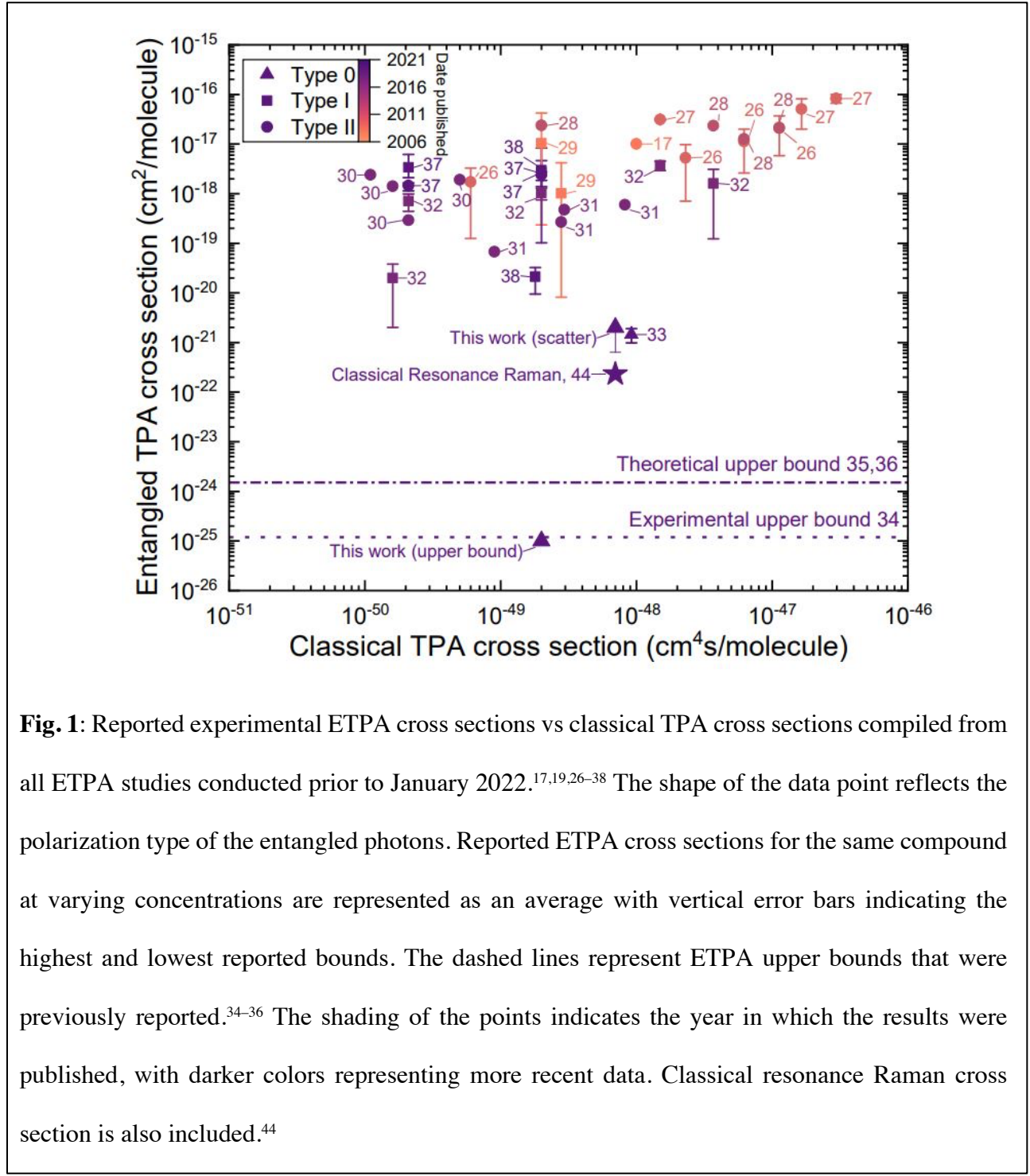

**Fig. 1**: Reported experimental ETPA cross sections vs classical TPA cross sections compiled from all ETPA studies conducted prior to January 2022.[17,19,26–38] The shape of the data point reflects the polarization type of the entangled photons. Reported ETPA cross sections for the same compound at varying concentrations are represented as an average with vertical error bars indicating the highest and lowest reported bounds. The dashed lines represent ETPA upper bounds that were previously reported.[34–36] The shading of the points indicates the year in which the results were published, with darker colors representing more recent data. Classical resonance Raman cross section is also included.[44]

Entangled photons are generated from a periodically poled 8% MgO doped congruent lithium tantalate (CLT) bulk crystal. The design and spectral characterization of the entangled photon source have previously been reported.[41] The crystal is optimized for type-0 energy-time entanglement by spontaneous parametric down-conversion (SPDC). A bandwidth of ~200 nm is produced by down-conversion of a 406 nm CW diode laser with a 1.5 nm linewidth. The large bandwidth and $10^{-8}$ conversion efficiency allow for a flux in the sub-µW range without exceeding the single-pair-per-mode quantum limit, as calculated by total photon pairs per second divided by the frequency bandwidth in Hz. The available photon flux is a significant improvement compared with the ~pW powers in most previous reports that use traditional bulk crystals such as beta barium borate (β-BBO).[17,26,31] The broad bandwidth also results in a ~20 fs correlation time (entanglement time) as measured after the sample. The entanglement time is not transform limited due to the group delay dispersion in the periodically poled chip and further optics. The high flux and tunable correlation times, from tens of femtoseconds to picoseconds, maximize the likelihood of measuring any entangled event.[35,36] The entangled photon bandwidth can be varied by tuning the temperature of the bulk crystal. For experiments in this paper, two distinct entangled photon bandwidths centered at 812 nm were utilized: a non-degenerate "split" SPDC spectrum spanning ~200 nm in wavelength and a degenerate "unsplit" SPDC spectrum (more details in the Supplementary Information). The two SPDC spectra are used to emphasize the scattering that can happen both close to and far from resonance of a molecule's absorption.

Rhodamine 6G was selected for this study because of its absorption near the SPDC pump laser wavelength,[39] its purely virtual two-photon classical absorption, as well as its high fluorescence quantum yield.[45] There are also several references of the proposed R6G ETPA from different measurement techniques.[33,34] A concentration of 5 mM R6G in ethanol is used for experiments; the classical single-photon fluorescence spectrum for this concentration is depicted in Fig. 4. Solvent effects and molecular dimerization at high sample concentration accounts for the redshifted spectrum.[46] Previously reported one-photon absorption cross sections for R6G in ethanol are on the order of $10^{-16}$–$10^{-17}$ cm$^2$/molecule and the two-photon absorption cross section is 70 (±10.5)×10$^{-50}$ cm$^4$s/molecule.[39] A higher 110 mM sample was also tested with the same conclusions. The 110 mM sample would have a 10-times higher fluorescence after the drop in quantum yield and self-absorption effects are taken into account, but the results of the paper were unchanged.

A custom two-photon Michelson-type interferometer, shown in Fig. 2, is used to characterize the entangled photon state. Specifically, the fourth-order interference, spectral characteristics, and entanglement (coherence) time of the broadband entangled photons are determined.[47,48] Type-0 SPDC was chosen because it utilizes the highest second-order nonlinear coefficient of LiTaO$_3$.[49] The two-photon Michelson-type interferometer allows the broad, collinear SPDC bandwidth to be more easily directed onto the sample. More specifically, the photon pairs are focused (beam waist 500 μm) onto the center of the cuvette. The measurement protocol is as follows: first, a cuvette containing pure ethanol is placed at the output of the Michelson interferometer. Second, the fourth-order interference is measured by collecting coincidences as a function of an optical time delay. The coincidence counting detection setup consists of a broadband 50:50 beamsplitter and two single photon avalanche photodiodes (SPADs, Laser Components). The fourth-order interference measurements are then repeated with the cuvette filled with R6G solution. Simultaneous to the coincidence counting measurement, a single photon counting EMICCD spectrometer (Princeton Instruments) is placed at 90 degrees relative to the excitation beam to measure any scatter or fluorescence. The collection efficiency including the spectrometer is ~10$^{-4}$, calculated from the measured classical single photon fluorescence and quantum yield of R6G. By monitoring the timescale and spectral features of the 90-degree signal, one can differentiate one-photon events, such as scattering, from an entangled two-photon absorption or fluorescence event.[33,34,50] A complete mathematical description of the interferometer output can be found in the Supplementary Information.

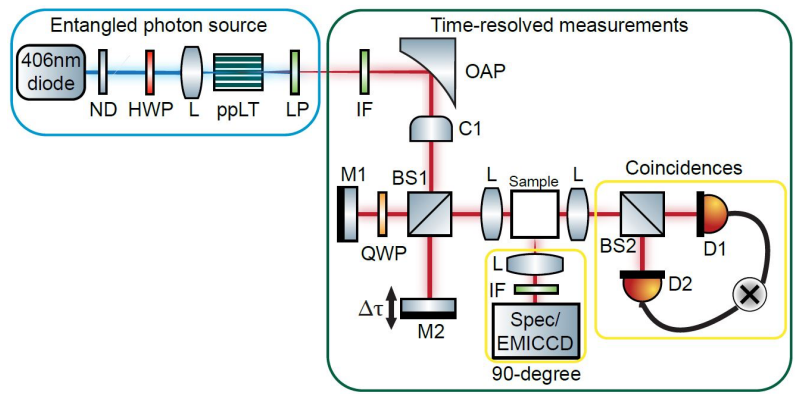

**Fig. 2:** Experimental setup used to measure the proposed entangled two-photon absorption and 90-degree signal. The entangled photon source consists of a continuous wave diode laser pumping a periodically poled lithium tantalate crystal. The resultant entangled photon cone is collimated before the two-photon Michelson interferometer. The fourth-order interference of the entangled photon pairs is measured with a coincidence counting configuration and the time-resolved 90-degree signal was imaged onto a spectrometer/EMICCD. ND: continuous neutral density filter wheel, HWP: half-wave plate, L: lens, ppLT: periodically poled lithium tantalate chip, LP: 500 nm longpass filter, IF: interference filter, OAP: off-axis parabolic mirror, C1: cylindrical lens, QWP: quarter-wave plate, BS1 and BS2: 50:50 plate beamsplitters, M1 and M2: mirrors, D1 and D2: multimode fiber coupled single photon avalanche diodes connected to a coincidence circuit, Spec/EMICCD: Spectrometer and electron multiplying intensified charge coupled device.

Figure 3 shows the fourth-order interference of the entangled photon state before and after interacting with 5mM R6G/ethanol as well as the signal collected at 90-degrees to the excitation pathway during the interaction. Corresponding theoretical fitting is included in the SI. The top trace is the measured coincidences through a cuvette filled with pure ethanol. In a two-photon Michelson interferogram, the degree of entanglement can be qualitatively inferred by examining the amplitude ratio above and below the oscillation baseline. A "top-heavy" feature, for example shown in Fig. 3 top trace, indicates photon bunching. Given the >95% visibility measured here and the fact that SPDC with a CW laser inherently creates an entangled state, we can reasonably confirm that our input source is entangled. For this experiment, the coherence/entanglement time is <20 fs and the interference visibility is 0.94. When the cuvette is filled with R6G (middle trace in Fig. 3), the visibility of the interference slightly decreases to 0.86, but the interferogram retains the spectral and temporal features shown with ethanol (with a slight shift in temporal patterns due to the change in refractive index. If ETPA does occur, then the photon bunching feature will diminish and the classical component will change spectrally due to the added fluorescence signal (or a nonuniform single photon absorption process).

To simulate fourth-order interferograms, the Fourier-transformed spectra of experimental interferograms are fitted to Gaussian peaks (SI Fig. S8). The fitted spectrum is then used as the input spectrum for an entangled two-photon Michelson simulation (SI Fig. S4). The bottom trace in Fig. 3 shows the summed spectrum collected at 90-degrees by the EMICCD. The interferogram follows that of a one-photon interference from a single Gaussian peak, also depicted in the SI Fig. S8. If the collected signal originated from an entangled two-photon event, it would follow the pattern of the input state and exhibit a matching two-photon interference as the molecule would act similar to a coincidence counter.[50]

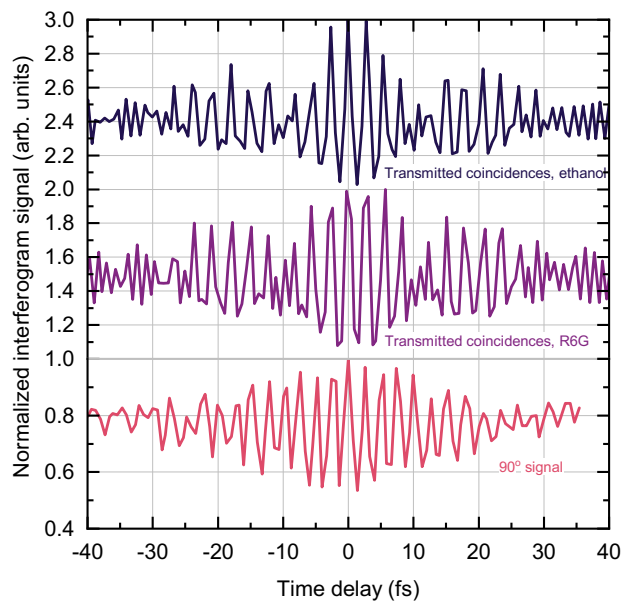

**Fig. 3**: Photon counts versus time delay for broadband entangled photons traveling through a two-photon Michelson interferometer. The top trace is transmitted coincidence counts through ethanol, the middle trace is transmitted coincidence counts through R6G in ethanol, and the bottom trace is summed EMICCD counts of the 90-degree signal. Comparison of these three signals indicates that only single photon events, and not an entangled two-photon event, is present for the R6G. The y axis has been scaled and shifted for visualization.

The one-photon nature of the scattered signal is further confirmed by looking at the wavelength spectrum collected at 90-degrees (Fig. 4). The measured single photon fluorescence of R6G is shown for reference in purple. For the non-degenerate entangled photon spectrum (Fig. S2), a scattered signal is measured that correlates with the wings of the spectrum, well away from the expected fluorescence spectrum. When a degenerate spectrum with a bandwidth centered at 800 nm is used, farther away from the R6G absorption tail, a reduced amplitude scatter at 800 nm is measured. The scattered spectrum that is measured is a convolution of the wavelength dependent molecular scattering cross section and the SPDC signal which is why it does not appear identical to Figure S2. As a control, zinc tetraphenylporphyrin (ZnTPP) was measured because it has a narrow-band absorption at 400 nm that is ~200 nm from the SPDC spectrum as well as a much weaker absorption near 550 nm (aligned with R6G). No scattered signal was measured for ZnTPP within the sensitivity limit of our spectrometer, which may be expected because ZnTPP has a lower absorption cross section than R6G, but other factors regarding the nature of the scattering process may come into play. All light is transmitted within our error bars and Figure S6 represents the dark count fluctuations of the EMICCD. At this point, the exact nature of the scattering mechanism is unknown, but the scattered spectrum and its single-photon time dependence are clearly measured.

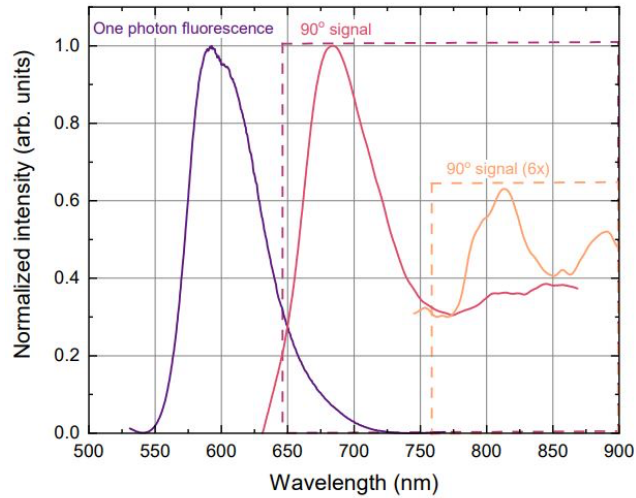

**Fig. 4**: One photon fluorescence (purple trace) and the measured scatter spectra from 5mM R6G in ethanol for non-degenerate (pink trace) and degenerate (orange trace) entangled photons. The spectra have been normalized and scaled such that they can be displayed on similar scales. The scattered signal matches the entangled photon spectrum indicating that a one-photon process is present, and not fluorescence from an entangled two-photon process.

The scattered cross section is calculated using the equation:[33]

$$S_E = \gamma\, N_E\, c\, *\, 10^{-6}\, *\, l N_A \sigma_E$$

where $S_E$ is the collected scatter signal, $\gamma$ is the collection efficiency, $N_E$ is the incident entangled photon flux, c is the concentration in mM, $l$ is the pathlength in cm, $N_A$ is Avogadro's number, and $\sigma_E$ is the entangled two-photon absorption cross section. The resulting scattered cross section is $2 \pm 1 * 10^{-21}$ cm²/molecule. For comparison, the measured attenuation from a transmission spectrum (Fig. 5), the standard method of the field, yields a cross section of $4 \pm 1 * 10^{-21}$ cm²/molecule cross section (See SI for more information on the measurement). The agreement between the measured scattered cross section and the measured attenuation cross section further confirms that no entanglement enhanced two-photon absorption or fluorescence is measured.

Our measurements indicate that off- or near-resonant scattering can lead to inflated absorption cross sections in intensity counting transmission, coincidence, or fluorescence experiments. The measured values are also above the maximum theoretical ETPA cross sections for R6G from quantum-mechanical derivations ($3 * 10^{-24}$ cm²/molecule, dashed line in Fig. 1).[36] The results are also in-line with the measurements of Parzuchowski et al,[34] where no signal was measured after extensive filtering and an ETPA cross section upper limit of $1.2 * 10^{-25}$ cm²/molecule was established for R6G. Note again that our spectrometer is capable of measuring cross sections down to the $10^{-25}$ cm²/molecule range, so if present, a fluorescence event would be detectable by these guidelines. The $10^{-25}$ cm²/molecule sensitivity is estimated based on the noise floor of the detector and the maximum possible photon count before detector saturation. At this sensitivity, we estimate the maximum fluorescence count rate, including collection efficiency, to be approximately 5-10 photons/s. The linear dependence measurement of Fig. 5 also covers input powers spanning four orders of magnitude, well beyond the intensity range reported in most manuscripts. The slope over this range is not completely linear but also not completely quadratic. If measured over a smaller range, the slope would appear linear, in agreement with previous reports. The scattered signal at 90 degrees, if not properly filtered, and measured sparsely, would also represent a signal that appears to match the entanglement time – emphasizing the importance of spectral *and* temporal measurements.

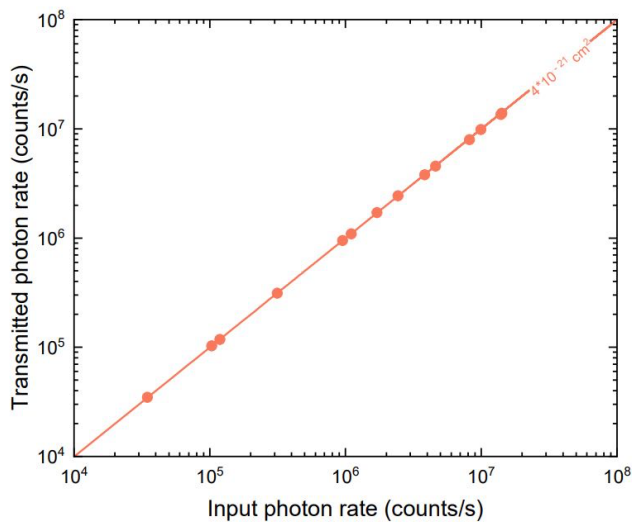

**Fig. 5:** Measured transmitted photon rate (orange data points) and fit (orange solid line) vs input photon rate for 5mM Rhodamine 6G in water. Using the method standard in literature, the ETPA cross section from the measured absorption rate would be $4\ (\pm 1) * 10^{-21}$ cm$^2$/molecule. In this manuscript, the linear relationship is determined to not be from ETPA, but rather from a single photon scattering process. Note that the error bars for each data point are on the order of the graph point size.

The measurements in this paper are performed on a virtual-state mediated two-photon excitable molecule. Recently, theoretical and experimental measurements have emphasized the importance of resonant or near resonant real states to achieve large two-photon entangled interactions.[42,43,51] The results of this paper indicate that caution must be taken when measuring entangled two-photon events because single-photon scattering and other reported single-photon processes such as hot band absorption[52] will mimic linear absorption at low power. Only by characterizing the entangled state, instead of relying solely on intensity measurements, can a definitive conclusion be reached. As the field of excited state entangled interactions continues to be explored, and molecules are synthesized to maximize entangled interactions, a standardized measurement scheme, such the one as presented here, is vital.

**EXPERIMENTAL METHODS**

*Experimental Configuration*

Broadband energy-time entangled photon pairs are generated via spontaneous parametric down-conversion of pump photons in a custom made periodically poled 8% MgO doped congruent lithium tantalate (CLT) bulk crystal (HC Photonics). The grating design is optimized for the Type-0 collinear quasi-phase matching condition and the poling period is 9.5 μm. The design and characterization of the grating has been previously reported.[41] The crystal is pumped by a continuous wave diode laser that outputs 400 mW of 406 nm light with a linewidth of 1.5 nm (Coherent OBIS). The polarization state of the diode laser is conditioned with a polarization beamsplitter and half-wave plate (Thorlabs) before the beam is focused into the crystal with a UV fused silica plano-convex lens. The temperature of the ppLT grating is maintained at phase-matching temperatures to an accuracy of 10 mK with a heater and PID loop (Covesion). After creation, the entangled photon pairs are passed through a total of three OD-4 500nm longpass filters (Edmund Optics) to eliminate any residual 406 nm pump photons.

For the static transmission and 90-degree signal measurements, the SPDC flux is focused into a cuvette containing Rhodamine 6G and the transmission or 90-degree signal is collected with a lens system to image with a spectrometer and an electron multiplying intensified CCD camera (PIMAX4, Princeton Instruments). The pump power is varied with a continuously variable ND wheel (Thorlabs) and monitored with a power meter (Newport). Additional longpass filters ensured that stray pump photons did not reach the EMICCD.

Time-resolved entangled state characterization was performed by first collimating the entangled photon pairs with an off-axis parabolic mirror and cylindrical lens. Following the collimation optics, the entangled pairs are directed into a two-photon Michelson interferometer.[47,48] This configuration is chosen over the Mach-Zehnder configuration as it enables a broader bandwidth of entangled photon pairs to be utilized in a simpler collinear geometry. A broadband, dispersion compensating beamsplitter (BS1, Layertec) probabilistically separates the entangled photon pairs and directs them to two sets of mirrors, which then reflect the photons such that they are spatially overlapped on the beamsplitter. In one arm, an achromatic quarter-wave plate (Thorlabs) rotates the polarization state of the entangled pairs. In the other arm, the optical path length is adjusted by scanning a mirror mounted on a translation stage (Newport), with a minimum 0.1 μm resolution that translates to a temporal resolution of 0.33 fs.

Coincidence counts are collected using a second broadband, dispersion compensating beamsplitter (Layertec) and two free-space-to-fiber setups that couple entangled photon pairs into single photon avalanche photodiodes (SPADs, Laser Component COUNT) connected to time-tagging electronics (PicoHarp 300). The free-space-to-fiber setups each consist of two mirrors to align the output paths of the second beamsplitter into a 4.51 mm focal length asphere (Thorlabs), which then focuses the SPDC into a multimode fiber (105 μm core diameter, 0.22 NA). A 500 nm longpass filter is placed before the asphere to ensure that stray pump light is not collected. For fourth-order interference measurements, the coincidence time-bin resolution is set to 8 ps and a coincidence histogram is collected at each stage position over an integration time of 5 seconds. Accidental coincidence counts are calculated by averaging the tail of each histogram and subtracting the average value from the entire histogram. The coincidence counts are then summed within a 14 ns window.

*Sample Preparation*

A volumetric pipette (Eppendorf) was used to add absolute ethanol (Sigma-Aldrich) to a vial containing powdered Rhodamine 6G (Sigma-Aldrich) to achieve desired concentrations.

**SUPPORTING INFORMATION**

The supporting information file contains:

- Additional experimental details for the linear absorption measurement procedure.

- Theoretical background for the signal measured by a two-photon Michelson interferometer and simulating interferometric signals.

- Additional data including entangled photon spectra, absorption spectra of studied molecular compounds, comparison of experimental and simulated interferograms for non-degenerate entangled photons, interferograms for degenerate entangled photons, 90-degree signal for ZnTPP, and destruction of interference with quarter-wave plate.

**AUTHOR INFORMATION**

**Notes**

**The authors declare no competing financial interests.**

**ACKNOWLEDGMENT**

This material is based upon work supported by the U.S. Department of Energy, Office of Science, Office of Basic Energy Sciences under Award Number(s) DE-SC0020151 (SKC).

Disclaimer: This report was prepared as an account of work sponsored by an agency of the United States Government. Neither the United States Government nor any agency thereof, nor any of their employees, makes any warranty, express or implied, or assumes any legal liability or responsibility for the accuracy, completeness, or usefulness of any information, apparatus, product, or process disclosed, or represents that its use would not infringe privately owned rights. Reference herein to any specific commercial product, process, or service by trade name, trademark, manufacturer, or otherwise does not necessarily constitute or imply its endorsement, recommendation, or favoring by the United States Government or any agency thereof. The views and opinions of authors expressed herein do not necessarily state or reflect those of the United States Government or any agency thereof.